# Programming Discrete Physical Systems

**Hermann von Issendorff**
Institut für Netzwerkprogrammierung
Hemmoor, Germany
hviss@issendorff.de

**Abstract.** Every algorithm which can be executed on a computer can at least in principle be realized in hardware, i.e. by a discrete physical system. The problem is that up to now there is no programming language by which physical systems can constructively be described. Such tool, however, is essential for the compact description and automatic production of complex systems. This paper introduces a programming language, called Akton-Algebra, which provides the foundation for the complete description of discrete physical systems. The approach originates from the finding that every discrete physical system reduces to a spatiotemporal topological network of nodes, if the functional and metric properties are deleted. A next finding is that there exists a homeomorphism between the topological network and a sequence of symbols representing a program by which the original nodal network can be reconstructed. Providing Akton-Algebra with functionality turns it into a flow-controlled general data processing language, which by introducing clock control and addressing can be further transformed into a classical programming language. Providing Akton-Algebra with metrics, i.e. the shape and size of the components, turns it into a novel hardware system construction language.

Comment: 24 pages, 18 figures

## 1. Introduction

The living nature demonstrates how to program discrete physical systems: It generates chains of amino acids from genetic code [1]. In a suitable environment the chain of amino acids then folds to a protein, i.e. a one-dimensional formation mutates into a three-dimensional one. The transaction can be reversed by changing the environment. This shows that there is a bijective and bicontinuous mapping between the chain of amino acids and the protein, called homeomorphism. With other words: The chain of amino acids represents a program which contains the complete spatial information of the protein.

The spatial structure of the protein arises from additional weaker binding forces between the amino acids. By external impact, e.g. by adsorption of another molecule, the structure of the protein may change into another metastable state. With other words: A protein has the ability to store and process information.

Pursuing this approach of nature, we present a many-sorted term algebra in this paper which is a programming language for the description of discrete spatiotemporal physical systems.

Discrete physical systems are a composition of a set of material components or any abstraction thereof. If the components are active, i.e. if they produce physical objects or evaluate functions, then they are temporally directed and are activated in a partial temporal order. If the components are static, then they can be assigned a partial assembling order which also induces a temporal direction. Abstracting a discrete physical system, for instance a computer, from its metrics, i.e. from the spatial measures of its components, the residue is a three-dimensional directed network of the executable functions which are realized by the components. Abstracting a discrete system, for instance again a computer, from its functionality, the residue is a three-dimensional directed network of building blocks which have the size and shape of the system components. Abstracting a discrete system from the metrics and the functionality at the same time, the residue is a three-dimensional network of nodes showing their dependencies. There, any two nodes may be related or not, and each node may be related to any finite number of preceding and succeeding nodes.

If the nodes of the network are provided again with their original concrete functional and metric properties, then the original system is regained. The nodal network is therefore a structural class of all discrete physical systems. Thus, if there is a formal language for the constructive description of a relational nodal network, it is a common language for all discrete physical systems. Akton-Algebra, for short *AA*, provides this capability.

It is important to notice that abstraction from metrics and functionality does not mean abstraction from space and time. The latter would reduce a directed discrete system to a directed graph, i.e. to a mathematical object of graph theory, which does not have any relation to space or time. The spatiotemporal relations between the nodes, however, are the very properties on which *AA* is based upon. An *AA*-node, i.e. the metric abstraction of a component, does still have a finite spatial extension, and traversing the node from its front-end to its back-end still takes a finite amount of time. This means that an action of a component does not disappear under metric and functional abstraction but is only reduced to a rudimentary action of propagation. This is the reason why the word "Akton" has been chosen as a general designator of a concrete component and any of its metrical or functional abstractions.

The constructive description of spatiotemporal structures by programming has not been recognized up to now. Classical data processing languages are not provided with spatial semantics. They do not need to because they are tailored to the sequential execution of the von-Neumann-computer, where the random-access-memory serves as a surrogate for space. At first glance, the graphical calculus proposed by Abramski and Coecke [2] seems to have some similarity to *AA*. Their calculus, however, is aimed at quantum informatics and does not describe classical physical structures. Thus, the only paper on spatiotemporal structures seems to be an early one by the author himself [3].

The paper proceeds as follows: In the next section the fundamental elements of a nodal network are analyzed. Additional elements need to be introduced in order to deal with crossings in spatial structures and cycles or crosslinks in planar structures. This gives rise to a hierarchy of akton sorts and a hierarchy of interface sorts. The language of abstract *AA* is then synthesized from these two hierarchies. There are three sets of production rules, one representing spatial by planar structures, another one representing planar by directed linear structures and a final one representing antiparallel directed linear structures. As demonstrated by some examples, abstract *AA* lends itself very well to modelling biological structures as for instance DNA, RNA and proteins. The rest of the paper is devoted to structural, functional and metrical concretizations. Structural specification can simply be achieved by adding more subsorts to the akton hierarchy. Functional specification requires the extension of both the akton and the interface hierarchies, and metrical specification also requires the extension of both hierarchies, although by different sorts. The conclusions finally subsume the major achievements of *AA*.

## 2. Elements and Elementary Structures of *AA*

As observed in the introduction, an abstract nodal network is a common structure underlying every discrete physical system. In general, a nodal network of a discrete physical system has a three-dimensional structure. A formal description, on the other hand, is an ordered sequence of symbols and thus has a one-dimensional structure. The aim of this chapter is to show that a nodal network can be mapped into a one-dimensional description and vice versa, without loosing any structural information.

The class of nodal networks we are dealing with is always assumed to be directed. If the physical system underlying the network is not directed or does not have an entry and an exit, these features can be introduced without the loss of generality. Each node of a directed nodal network has two interfaces, one for the input and one for the output.

The representation of a directed abstract nodal network can be done at different levels of detail. As depicted by the hierarchy in figure 1, new levels of sorts with more and more properties can be added. At the top level, the network is represented by the general sort *Akton*. At the second level, called the fundamental level, there are four sorts of nodes called *Head, Body, Tail* and *CS* (*Closed System*). The relations between the sorts are specified by means of interfaces. At the fundamental level, there are only two primitive sorts of interfaces, the non-empty one (symbolized by $\neg \varepsilon$) and the empty one (symbolized by $\varepsilon$). Sort *Head* has no preceding nodes, i.e. an empty input but at least one succeeding node, i.e. a non-empty output. Sort *Tail* has no succeeding nodes but at least one preceding node, i.e. a non-empty input and an empty output. Sort *Body* has at least one preceding node and at least one succeeding node, i.e. a non-empty input and a non-empty output. Finally, sort *CS* has no preceding and no succeeding nodes, i.e. input and output are empty.

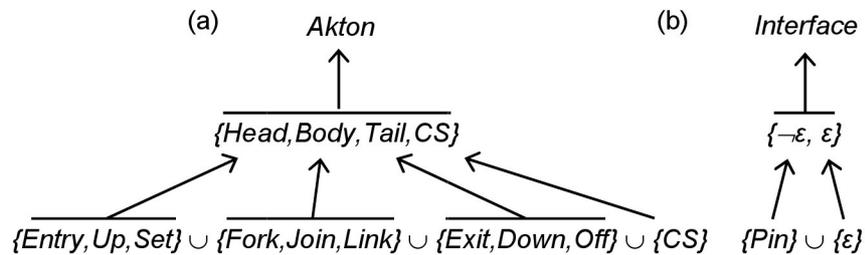

Figure 1. (a) The hierarchy of abstract akton sorts. These three levels are common to all discrete systems. The third level comprises the set of basic abstract sorts. (b) The hierarchy of interface sorts is depicted at the right. At the third level, the non-empty interface sort is quantified by a basic element called Pin.

The third level, called abstract structural level, consists of basic abstract sorts. Their non-empty inputs and outputs are specified by a quantified basic interface element called *Pin*. There are three basic subsorts of *Body* called *Fork*, *Join* and *Link*. *Fork* has one *Pin* in the input and two *Pins* in the output, *Join* has two *Pins* in the input and one *Pin* in the output, and *Link* has one *Pin* in the input and one in the output. The basic subsorts of *Head* are called *Entry*, *Up* and *Set* each of them having a single *Pin* in the output. The basic subsorts of *Tail* are called *Exit*, *Down* and *Off*, each having a single *Pin* in the input. The subsorts of *Head* and *Tail* differ by their semantics.

While *Entry* and *Exit* represent the boundaries to the outside world, *Up* and *Down* as well as *Set* and *Off* are necessary for mapping the three dimensions of nodal networks to the single dimension of a string of symbols, i.e. the program code. This mapping needs to be explained more closely.

Because of the abstraction from metrics the structure of a nodal network is a topological one. A topological structure preserves the adjacency of the nodes, if it is mapped into another shape by a function called homeomorphism. Even cuts are admissible as long as the correlation between the cutting ends is guaranteed. Homeomorphism is bijective and bicontinuous. This means that an original topological structure may arbitrarily be distorted but can always be regained by reversing the homeomorphism.

In order to describe the topological properties of the nodal network, a topological frame of reference is needed. Since there is no metrics, the frame of reference can only be relational. Such frame of reference can be defined by referring to a human observer who physically differentiates between three independent pairs of inverse spatial relations, i.e. *left-right*, *above-below* and *front-back*.

In addition, a privileged direction in the frame of reference needs to be selected in order to orient the directed nodal system, i.e. on which side to place the elements of sort *Head* and on which side the elements of sort *Tail*. Following the reading standard of the Western Hemisphere, the direction from *left* to *right* is chosen. Since every node represents an action and action takes time, the orientation also introduces a direction of time. Thus *left* will also be interpreted as *earlier* and *right* as *later*.

The mapping of the nodal network from a three-dimensional representation to the one-dimensional description of *AA* requires several steps:

1. The nodal network has to be oriented so that all system *Entries* are at the left side and all system *Exits* are at the right side.

2. The nodal network has to be projected to an oriented plane of observation spanned between the *left-right* axis and the *above-below* axis and positioned between the nodal network and the observer. Usually, this projection will give rise to crossings of nodal connections (see figure 2). Since the crossings are spatial residues they must be removed. This is achieved by cutting the lower connection and replacing the cutting ends by a pair of *Down* and *Up, Down* being a subsort of *Tail* and *Up* being a subsort of *Head* as mentioned before.

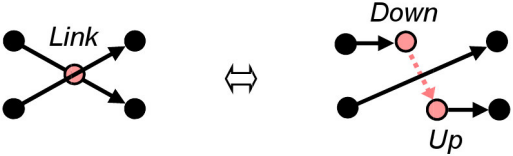

Figure 2: Planarization of a spatial structure, i.e. removal of a crossing, is achieved by cutting the lower connection and inserting a pair of aktons called Down and Up.

3. The resulting planar network may still contain non-orientable structures like cycles or crosslinks (see figure 3). This problem can be solved in a way similar to the crossing problem by cutting the structures and inserting a pair *Off* and *Set,* representing a cut in the plane. *Off* is a subsort of *Tail* and *Set* is a subsort of *Head* as mentioned before.

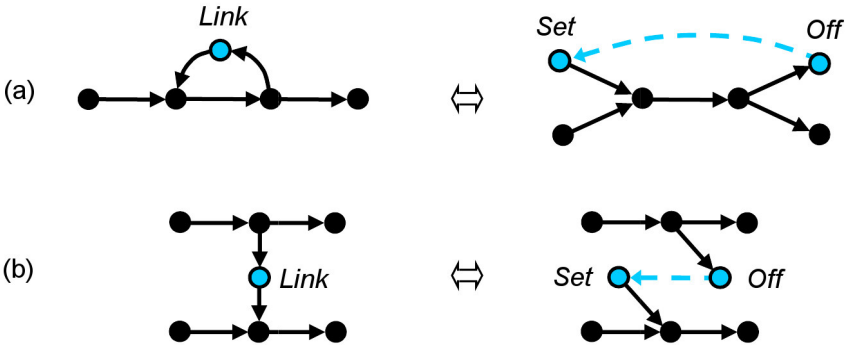

Figure 3: Orientation of cycles (a) and crosslinks (b) from left to right by cutting and inserting a pair of aktons called *Off* and *Set*.

4. The separate utilization of spatial and planar cuts does not suffice to linearize every spatial nodal network. There are nodal structures where both cuts are to be applied crosswise. These structures can be characterized by two antisense strands which are interconnected by several crosslinks. A simplified structure of this kind is depicted in figure 4 showing two antisense strands which are connected by two crosslinks. The directions of these crosslinks are not specified on purpose. An orientation of them can be accomplished by either left- or right-twisting the contrarily oriented strand thus generating different crossings of the crosslinks.

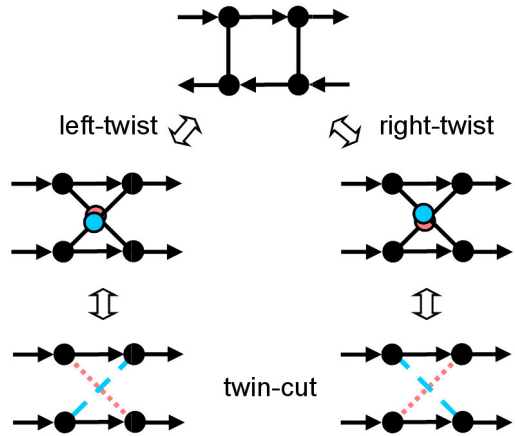

Figure 4: Simplified structure of two antisense strands connected by two undirected crosslinks.

The planarization of the crossings is achieved by applying a spatial cut to the lower crosslink as shown by the dotted red line, and a planar cut to the upper one as shown by the dashed blue line. Finally assigning directions to the crosslinks amounts to four different twin-cuts for the left-twisted as well as the right-twisted structure. The twin-cuts of the left-twisted structures are depicted in detail in figure 5.

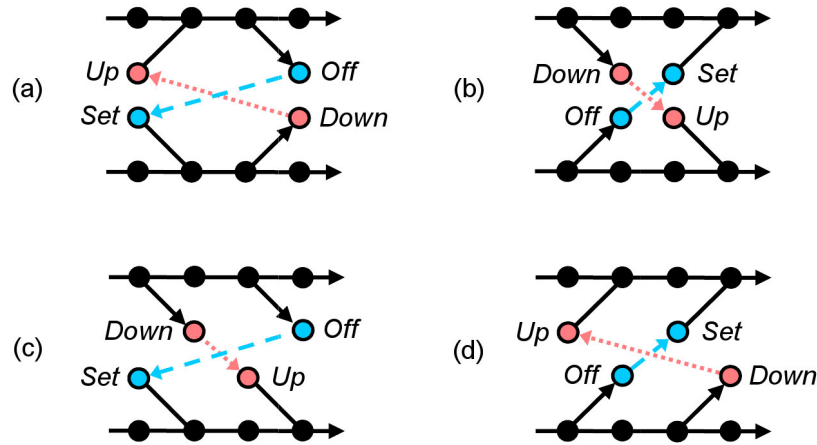

Figure 5: Detailed view of four left-twisted twin-cut structures (a), (b), (c), (d) which can be derived from the undirected twin-cut structure of figure 4.

A nodal network can now formally be represented by a string of symbols, i.e. in linear form. To this end, two adjacent independent subnetworks $x$ and $y$ are related by an infix symbol "/", called *Juxta*, where *x/y* means $x$ lies *above* $y$. Likewise, two adjacent dependent subnetworks $x$ and $y$ are related by an infix symbol ">", called *Next*, where *x>y* means *x precedes y* in space and time. In order to reduce the amount of parentheses *Juxta* is assumed to bind stronger than *Next*.

## 3. Definition of Abstract AA

In the previous chapter we discovered a way how to turn a three-dimensional network of nodes into a linear oriented network of a nodal string, i.e. a string of abstract aktons. A nodal string can be read and processed like a program, which under abstraction from functionality and metrics means to reconstruct the original spatial topological structure of the nodal network. Thus we already know that there exists a formal language for the description of nodal networks. In particular, we discovered the set of basic structural nodes of *AA*. However, what we cannot do up to now is to synthesize a nodal system, because we do not know the rules by which nodal subnetworks are to be related. In formal language terms, we need to know the grammar of *AA*. That is what we will do in this chapter.

*AA* is a many-sorted term-algebra. It is built up from a hierarchy of akton sorts and a hierarchy of akton interfaces.

The grammar of *AA* is systematically derived descending the hierarchy of sorts as shown in figure 1. The hierarchy is headed by sort *Akton*. This general notion will step by step be specified while descending the hierarchy further down.

Aktons and the relations *Next* and *Juxta* are destined to describe directed nodal networks consisting of inner nodes, i.e. of sort *Body*, and of outer nodes, i.e. of sorts *Head* and *Tail*. In addition, there is the akton sort *CS* describing a complete network. The four sorts establish a set *FA* of fundamental akton sorts.

**Definition A.1 (Fundamental Akton Sorts)**
*The set FA of fundamental akton sorts is defined as*
*FA := {Head, Body, Tail, CS}.*

Adjacent directed aktons may either be dependent or independent. Dependency is expressed by the relation *Next* and independency by the relation *Juxta*. *Next* and *Juxta* produce a free semigroup $FA^+$ of fundamental akton terms.

The term variables defined here and further down by means of tables are represented by $z \in z^+$, i.e. by using the proper names of the term sorts. The representation of the production rules by tables is nothing but an inverse BNF.

**Definition A.2 (Fundamental Akton Terms)**
*The set of fundamental akton terms $FA^+$ is inductively defined as the smallest set satisfying:*
- $FA^+ := H^+ \cup T^+ \cup B^+ \cup CS^+$,
- $Head \in H^+, Body \in B^+, Tail \in T^+, CS \in CS^+$,
- $\forall x,y \in FA^+$:

$(x/y) \in FA^+$, if          $(x>y) \in FA^+$, if

|       | y=  |   |   |    |
|-------|-----|---|---|----|
| (x/y) | H   | B | T | CS |
| H     | H   | B | B | H  |
| B     | B   | B | B | B  |
| T     | B   | B | T | T  |
| CS    | H   | B | T | CS |

x=

|       | y= |    |
|-------|----|----|
| (x>y) | B  | T  |
| H     | H  | CS |
| B     | B  | T  |

x=

The nodal structures presented by aktons of sort *Body* are subdivided into three subsorts which are called *Fork, Join and Link* accordingly. As discussed in the previous chapter, the mapping of spatial structures to planar structures requires a cut, represented by the pair of akton sorts *Down* and *Up*. Likewise, the complete sequencing of planar structures requires another pair of akton sorts *Off* and *Set*. Thus, *Head* comprises the subsorts *Entry, Up* and *Set* and *Tail* the subsorts *Exit, Down* and *Off*.

**Definition A.3 (Structural Akton Sorts)**
*The set A of structural akton sorts is defined as*
*A := Head $\cup$ Tail $\cup$ Body $\cup$ {CS}, where*
*Head := {Entry, Up, Set},*
*Tail := {Exit, Down, Off},*
*Body := {Fork, Join, Link}.*

Every akton term *x* has two interfaces, an input and an output, formally described by *in(x)* and *out(x)*. Depending on the level of concretization, the interfaces may be refined by introducing more properties. This gives rise to a hierarchy of interface sorts, similar to the hierarchy of akton sorts. At the third level of the specification hierarchy (see figure 1) an interface consists of a set of ordered elements of sort *Pin* or may be empty. As usual, the symbol "$\varepsilon$" stands for empty.

**Definition A.4 (Interface Sorts)**
*The set I of interface sorts is defined as I := Pin $\cup$ $\varepsilon$.*

The interfaces of *Juxta*-related akton terms are packed on top of each other, in the same way as the akton terms themselves. Accordingly, the *Juxta*-symbol is also used for the interface elements. The order of the interface elements matches the structural order of adjacent dependent nodes, i.e., the order serves as a physical addressing scheme. The set of interface terms $I^*$ is associative.

**Definition A.5 (Interface Terms)**
*The set of interface terms $I^*$ is inductively defined as the smallest set satisfying:*
- $I \subset I^*$
- $\forall i,j,k \in I^*$: $(i/j) \in I^*$ and $(i/j)/k = i/(j/k)$ and $i/\varepsilon = \varepsilon/i = i$

The functions *in* and *out* retrieve the input respectively the output of an akton term. Both functions serve to define the interfaces of the aktons of set *A*.

**Definition A.6 (Functions *in*, *out*)**
*The functions in, out: $A^+ \rightarrow I^*$ are inductively defined as:*

$$in(z) := \begin{cases} \varepsilon, & \text{if } z \in Head \cup CS \\ Pin, & \text{if } z \in Tail \cup Link \\ Pin, & \text{if } z = Fork \\ Pin/Pin, & \text{if } z = Join \\ in(x)/in(y), & \text{if } z = (x/y) \\ in(x), & \text{if } z = (x>y) \end{cases} \quad out(z) := \begin{cases} \varepsilon, & \text{if } z \in Tail \cup CS \\ Pin, & \text{if } z \in Head \cup Link \\ Pin/Pin, & \text{if } z = Fork \\ Pin, & \text{if } z = Join \\ out(x)/out(y), & \text{if } z = (x/y) \\ out(y), & \text{if } z = (x>y) \end{cases}$$

The set of akton terms $A^+$ is a subset of $FA^+$. A first restriction of $A^+$ is that two dependent terms can only be *Next*-related if their adjacent interfaces are identical. The other restriction is that every relation of terms must conform to a real topological structure.

A crossing for instance is characterized by the cut terms *D*, $D \in D^+$, and *U*, $U \in U^+$, which are separated by a term *B*, $B \in B^+$. This gives rise to four special term sorts, designated by $DB^+$, $BD^+$, $UB^+$, $BU^+$ (see figure 2). There are four of them because every crossing owns a chirality that can be either left- or right-handed.

The planar structures of cycles and crosslinks on the other hand are characterized by the cut terms *O*, $O \in O^+$, and *S*, $S \in S^+$, which are not separated by another term (see figure 3). A pair of cut terms can thus only be positioned either at the left or at the right side of a *B*-term, which results in four special term sorts $OB^+$, $BO^+$, $SB^+$, $BS^+$. Four more special term sorts $OBO^+$, $OBS^+$, $SBO^+$, $SBS^+$ need to be introduced because independent cut terms may occur at both sides of a *B*-term.

The set of production rules $P$ consists of four subsets $P_i$, $i \in \{1,2,3,4\}$. $P_1$ lists the spatial rules, $P_2$ the planarizing rules, $P_3$ the linearizing rules and $P_4$ the twin-cut rules.

**Definition A.7 (Terms of the Akton Language)**
*The set of akton terms $A^+$ is inductively defined as the smallest set satisfying:*
- $A^+ \subset FA^+$
- $A \subset A^+$
- $A^+ := B^+ \cup E^+ \cup X^+ \cup CS^+ \cup U^+ \cup D^+ \cup S^+ \cup O^+$
- $Body \in B^+$, $Entry \in E^+$, $Exit \in X^+$, $CS \in CS^+$, $Up \in U^+$, $Down \in D^+$, $Set \in S^+$, $Off \in O^+$
- $\forall x,y \in A^+:$ $(x>y) \in A^+$, if $out(x)=in(y)$ and $(x>y) \in P$
- $\forall x,y \in A^+:$ $(x/y) \in A^+$, if $(x/y) \in P$
- $P := P_1 \cup P_2 \cup P_3 \cup P_4$

**Definition A.8 (Spatial Production Rules of the Akton Language)**
*The set of spatial production rules $P_1$ are defined as:*

$x=$

| (x/y) | E | B | X | CS |
|---|---|---|---|---|
| E | E | B | B | E |
| B | B | B | B | B |
| X | B | B | X | X |
| CS | E | B | X | CS |

$y=$

$x=$

| (x>y) | B | X |
|---|---|---|
| E | E | CS |
| B | B | X |

$y=$

**Definition A.9 (Planarizing Production Rules of the Akton Language)**
*The set of planarizing production rules $P_2$ are defined as:*

$x=$

| (x/y) | B | U | D | BU | UB | DB | BD | CS |
|---|---|---|---|---|---|---|---|---|
| B | B | BU | BD | BU |  |  | BD | B |
| U | UB | U |  |  | UB |  |  | U |
| D | DB |  | D |  |  | DB |  | D |
| BU |  | BU |  | BU |  |  |  | BU |
| UB | UB |  |  |  | UB |  |  | UB |
| DB | DB |  |  |  |  | DB |  | DB |
| BD |  |  | BD |  |  |  | BD | BD |
| CS | B | U | D | BU | UB | DB | BD | CS |

$y=$

$x=$

| (x>y) | D | BU | UB | BD | DB |
|---|---|---|---|---|---|
| U | CS |  |  |  |  |
| DB |  | B |  |  |  |
| BD |  |  | B |  |  |
| UB |  |  |  | B |  |
| BU |  |  |  |  | B |

$y=$

The complete definition of crossings (see figure 2) requires the additional definition of the implicit *in-out*-relations between *U*- and *D*-terms. The two definitions regard the mirrored structures alternatively crossing a *B*-term either from above or below.

**Definition A.10 (Implicit Cut-Relations in Crossings)**
*The implicit cut-relations in crossings are defined as:*
- $\forall v,w,x,y \in A^+$: $out(y) := in(v)$, if $(v/w>x/y) \in B^+$ and $y \in U^+$ and $v \in D^+$ or
  $out(x) := in(w)$, if $(v/w>x/y) \in B^+$ and $x \in U^+$ and $w \in D^+$.

**Definition A.11 (Linearizing Production Rules of the Akton Language)**
*The set of linearizing production rules $P_3$ are defined as:*

|       | (x/y) | B  | S   | O   | BO  | BS  | OB  | SB  | CS |
|-------|-------|----|-----|-----|-----|-----|-----|-----|----|
|       | B     | B  | BS  | BO  | BO  | BS  |     |     | B  |
|       | O     | OB | B   | O   | OBO | OBS |     |     | O  |
|       | S     | SB | S   | B   | SBO | SBS |     |     | S  |
| x=    | OB    | OB | OBS | OBO | OBO | OBS |     |     | OB |
|       | SB    | SB | SBS | SBO | SBO | SBS |     |     | SB |
|       | BS    |    |     |     |     |     | B   |     | BS |
|       | BO    |    |     |     |     |     |     | B   | BO |
|       | CS    | B  | S   | O   | BO  | BS  | OB  | SB  | CS |

(header row: y=)

|       | (x>y) | O   | B   | BO | OB | BS | SB | OBO | OBS | SBO | SBS |
|-------|-------|-----|-----|----|----|----|----|-----|-----|-----|-----|
|       | S     | CS  | S   |    |    |    |    |     |     |     |     |
|       | B     | O   | B   | BO | OB | BS | SB | OBO | OBS | SBO | SBS |
|       | BS    |     | BS  | B  |    |    |    | OB  |     | SB  |     |
|       | SB    |     | SB  |    | B  |    |    | BO  | BS  |     |     |
| x=    | BO    |     | BO  |    |    | B  |    |     | OB  |     | SB  |
|       | OB    |     | OB  |    |    |    | B  |     |     | BO  | BS  |
|       | SBS   |     | SBS | SB | BS |    |    | B   |     |     |     |
|       | SBO   |     | SBO |    | BO | SB |    |     | B   |     |     |
|       | OBS   |     | OBS | OB |    |    | BS |     |     | B   |     |
|       | OBO   |     | OBO |    |    | OB | BO |     |     |     | B   |

(header row: y=)

The complete definition of cycles and crosslinks requires the additional definition of the implicit *in-out* relations between *S-* and *O-*terms. Recall that the feedback of a cycle is always located above or below a *B-*term (see figure 3a). A crosslink, on the other hand, is located between two *B-*terms connecting an upper *B-*term with a lower *B-*term or vice versa (see figure 3b).

**Definition A.12 (Implicit Cut-Relations in Cycles)**
*The implicit cut-relations in cycles are defined as:*
- $\forall v,w,x,y \in A^+$: $out(v) := in(x)$, if $(v/w>x/y) \in B^+$ and $v \in S^+$ and $x \in O^+$ or
  $out(w) := in(y)$, if $(v/w>x/y) \in B^+$ and $w \in S^+$ and $y \in O^+$.

**Definition A.13 (Implicit Cut-Relations in Crosslinks)**
*The implicit cut-relations in crosslinks are defined as:*
- $\forall u,v,w,x,y,z \in A^+$: $out(v) := in(y)$, if $(u>x/y)/(v/w>z) \in B^+$ and $v \in S^+$ and $y \in O^+$ or
  $out(v) := in(y)$, if $(u/v>x)/(w>y/z) \in B^+$ and $v \in S^+$ and $y \in O^+$.

The final structure to be defined by production rules is the twin-cut structures as outlined in section 4 of the previous chapter. As mentioned the twin-cut structures may be either left- or right-twisted. Changing the direction of the spatial as well as the planar cut gives rise to four different structures each. The left-twisted structures are depicted in figure 5.

**Definition A.14 (Twin-Cut Production Rules of the Akton Language)**
*The set of the twin-cut production rules $P_4$ are defined as:*

<table>
<tr><td colspan="6" align="center">*left-twisted*</td></tr>
</table>

|       | (x>y) | BO   | BS   | DB   | UB   |
|-------|-------|------|------|------|------|
|       | BU    | BUBO | BUBS |      |      |
| x=    | BD    | BDBO | BDBS |      |      |
|       | SB    |      |      | SBDB | SBUB |
|       | OB    |      |      | OBDB | OBUB |

|       | (x>y) | BU   | BD   | SB   | OB   |
|-------|-------|------|------|------|------|
|       | BO    | BOBU | BOBD |      |      |
| x=    | BS    | BSBU | BSBD |      |      |
|       | DB    |      |      | DBSB | DBOB |
|       | UB    |      |      | UBSB | UBOB |

|       | (x/y) | SBDB | OBUB | SBUB | OBDB |
|-------|-------|------|------|------|------|
|       | BUBO  | B    |      |      |      |
| x=    | BDBS  |      | B    |      |      |
|       | BDBO  |      |      | B    |      |
|       | BUBS  |      |      |      | B    |

|       | (x/y) | DBSB | UBOB | UBSB | DBOB |
|-------|-------|------|------|------|------|
|       | BOBU  | B    |      |      |      |
| x=    | BSBD  |      | B    |      |      |
|       | BOBD  |      |      | B    |      |
|       | BSBU  |      |      |      | B    |

**Definition A.15 (Implicit Twin-Cut Relations in Left-twisted Antiparallel Structures)**
*The implicit cut-relations in left-twisted* antiparallel structures *are defined as:*
- $\forall s,t,u,v,w,x,y,z \in A^+$: *out(t):=in(y) and out(u):=in(x), if (s/t>w/x)/(u/v>y/z)$\in B^+$ and*
  $t \in U^+$ *and* $y \in D^+$ *and* $x \in O^+$ *and* $u \in S^+$,
- $\forall s,t,u,v,w,x,y,z \in A^+$: *out(x):=in(u) and out(y):=in(t), if (s/t>w/x)/(u/v>y/z)$\in B^+$ and*
  $y \in U^+$ *and* $t \in D^+$ *and* $u \in O^+$ *and* $x \in S^+$,
- $\forall s,t,u,v,w,x,y,z \in A^+$: *out(y):=in(t) and out(u):=in(x), if (s/t>w/x)/(u/v>y/z)$\in B^+$ and*
  $y \in U^+$ *and* $t \in D^+$ *and* $x \in O^+$ *and* $u \in S^+$,
- $\forall s,t,u,v,w,x,y,z \in A^+$: *out(t):=in(y) and out(x):=in(u), if (s/t>w/x)/(u/v>y/z)$\in B^+$ and*
  $t \in U^+$ *and* $y \in D^+$ *and* $u \in O^+$ *and* $x \in S^+$.

**Definition A.16 (Implicit Twin-Cut Relations in Right-twisted Antiparallel Structures)**
*The implicit cut-relations in right-twisted* antiparallel structures *are defined as:*
- $\forall s,t,u,v,w,x,y,z \in A^+$: *out(t):=in(y) and out(u):=in(x), if (s/t>w/x)/(u/v>y/z)$\in B^+$*
  *and* $u \in U^+$ *and* $x \in D^+$ *and* $y \in O^+$ *and* $t \in S^+$,
- $\forall s,t,u,v,w,x,y,z \in A^+$: *out(x):=in(u) and out(y):=in(t), if (s/t>w/x)/(u/v>y/z)$\in B^+$*
  *and* $x \in U^+$ *and* $u \in D^+$ *and* $t \in O^+$ *and* $y \in S^+$,
- $\forall s,t,u,v,w,x,y,z \in A^+$: *out(y):=in(t) and out(u):=in(x), if (s/t>w/x)/(u/v>y/z)$\in B^+$*
  *and* $x \in U^+$ *and* $u \in D^+$ *and* $y \in O^+$ *and* $t \in S^+$,
- $\forall s,t,u,v,w,x,y,z \in A^+$: *out(t):=in(y) and out(x):=in(u), if (s/t>w/x)/(u/v>y/z)$\in B^+$*
  *and* $u \in U^+$ *and* $x \in D^+$ *and* $t \in O^+$ *and* $y \in S^+$.

Single *Forks*, *Joins* and *Links* have a planar structure, and multiple *Links* are also planar. The structures of *multiple Forks* and *multiple Joins*, however, are always spatial, i.e. projecting them on a plane can only be achieved by means of *Down/Up*-cuts. A *multiple Fork* structure duplicates a structure of *multiple Links* into two structures of *multiple Links*, and a *multiple Join* term does the reverse. In order to formally describe these structures, we need to introduce the separation functions *pre* and *suc*, which split a *Next*-relation into a preceding and a succeeding part.

**Definition A.17 (Separation Functions *pre*, *suc*)**
*The functions pre, suc: $A^+ \to A^+$ are defined as:*
*pre(x>y): = x,*
*suc(x>y): = y.*

**Definition A.18 (Multiple Link)**
*The set of link terms $L^+$ is inductively defined as the smallest set satisfying:*
- $L^+ \subset B^+$
- $Link \in L^+$
- $\forall x \in L^+: Link/x \in L^+$

M*ultiple Forks* as well as a *multiple Joins* are inherently asymmetric because they can be realized by a *Down/Link* term as well as by a *Link/Down* term. Accordingly, there are two sets of *multiple Forks* and of *multiple Joins*. They are distinguished by the subscripts *l* and *r* which indicate whether the Down is at the left side or right side of the Link.

**Definition A.19 (Multiple Fork)**
*The sets of multiple fork terms are recursively defined as:*
- $F_l^+ \cup F_r^+ \subset B^+$,
- $(Fork{>}Down/Link) > (Link/Up) \in F_l^+$,
- $\forall x \in F_l^+: (pre(x)/(Fork{>}Down/Link) > Link/suc(x)/Up) \in F_l^+$
- $(Fork{>}Link/Down) > (Up/Link) \in F_r^+$,
- $\forall x \in F_r^+: (pre(x)/(Fork{>}Link/Down) > Up/suc(x)/Link) \in F_r^+$.

**Definition A.20 (Multiple Join)**
*The sets of multiple join terms are recursively defined as:*
- $J_l^+ \cup J_r^+ \subset B^+$,
- $(Down/Link) > (Link/Up{>}Join) \in J_l^+$,
- $\forall x \in J_l^+: (Down/pre(x)/Link > suc(x)/(Link/Up{>}Join)) \in J_l^+$.
- $(Link/Down) > (Up/Link{>}Join) \in J_r^+$,
- $\forall x \in J_r^+: (Link/pre(x)/Down > suc(x)/(Up/Link{>}Join)) \in J_r^+$.

Modularity, i.e. the capability of combining several modules into a single one, is an indispensable requirement for the design of complex systems. In *AA* modularity is easily incorporated because all modules are aktons, and every akton term, how big it ever may be, can be concealed into a single akton. Concealing means hiding the structure of an akton term into an akton while preserving the visibility of the input and the output. The new akton is added to set *A* and of course needs to be provided with a distinct name. In contrast, regarding conventional digital programming languages, modularization and information hiding can be quite a problem [4].

**Definition A.21 (Function *conceal*)**
*The function conceal: $A^+ \rightarrow A$ is defined as:*
*conceal(x) := a, if not $x \in A$*

In order to simplify the akton expressions we introduce a count function for sequences of identical *Next*-terms and a count function for regular *Juxta*-terms. *Next*-counts are represented by a \*-symbol and *Juxta*-counts by a ^-symbol. Syntactically, both symbols bind stronger than *Next* and *Juxta*.

**Definition A.22 (Counting Functions \* and ^)**
*The counting function \*: $N \times A^+ \rightarrow A^+$ is inductively defined as:*
- $n*x = x{>}(n-1)*x,\ 0*x{>}y = y{>}0*x = y$

*The counting function ^: $A^+ \times N \rightarrow A^+$ is inductively defined as:*
- $x^{\wedge}n = x/x^{\wedge}(n-1),\ y/x^{\wedge}0 = x^{\wedge}0/y = y$

## 4. Dependency Preserving Term Replacements

The structure of a given nodal network can be modified in different ways without affecting the dependencies between the terms. Formally the modifications are achieved by term replacement according to the rules of Tab. 1. The rules say that the left term may be replaced by the right term provided that the constraint at the right side holds. The "↔"-symbol says that the terms are mutually replaceable.

The link-rules a. add a neutral term *y* to a term *x* or delete it. The neutral term *y* may either precede or succeed term *x* as stated by the two rules. The constraints are that the output of the left term must fit the input of the right term. Usually term *y* will just consist of a strip of *Links*. The expansion-rules b. place or remove a dead term *y*, i.e. a neutral place, above or below a term *x*. Term *x* is of sort $A^+$, term *y* is of sort $CS^+$. Both expansion-rules play an important role in the layout process. The associativity-rules c. modify the structure of *Next-* and *Juxta*-related terms. The first of the distributivity-rules d. states that distributivity of *Juxta* over *Next* always holds while the next rule states that distributivity of *Next* over *Juxta* is restricted. The connectivity-rules e. splice two independent *Juxta*-terms into a single term or vice versa.

**Table 1:** Dependency preserving term replacement rules

| | | |
|---|---|---|
| a. Link-Rules: | $x \leftrightarrow (y>x)$, | if $in(y)=in(x)$ |
| | $x \leftrightarrow (x>y)$, | if $out(y)=out(x)$ |
| b. Expansion-Rules: | $x \leftrightarrow (y/x)$, | if $y \in CS^+$ |
| | $x \leftrightarrow (x/y)$, | if $y \in CS^+$ |
| c. Associativity-Rules: $((x>y)>z) \leftrightarrow (x>(y>z))$, | | if true |
| $((x/y)/z) \leftrightarrow (x/(y/z))$, | | if true |
| d. Distributivity-Rules: $((w>x)/(y>z)) \rightarrow (w/y>x/z)$, | | if true |
| $(w/y>x/z) \rightarrow ((w>x)/(y>z))$, | | if $out(w)=in(x)$ |
| e. Connectivity-Rules: $((w>x)/(y>z)) \leftrightarrow (w>x/y>z)$, | | if $out(x)=\varepsilon$ and $in(y)=\varepsilon$ |
| $((w>x)/(y>z)) \leftrightarrow (y>w/z>x)$, | | if $in(w)=\varepsilon$ and $out(z)=\varepsilon$ |
| | | $w,x,y,z \in A^+$ |

## 5. Abstract structural models

The properties of abstract *AA* are exemplified by four nodal structures and their description by an *AA*-program.

### 5.1 Tetrahedron
The first example (see figure 6) deals with a tetrahedron showing in detail the successive steps of mapping from the spatial structure to the linear program. In a first step a rear edge of the tetrahedron is cut, as marked by a thin red line, and the free ends are marked by an *Up/Down*-pair. This makes it possible to spread the tetrahedron on a plane, and to orient the planar structure from left to right according to the direction introduced by the *Up/Down*-pair. The planar structure is then cut again, as marked by a long thin blue line. While the cuts of both outer edges are healed by inserting *Links* the cut of the crosslink is marked by a *Set/Off*-pair, all shown in blue. This provides the crosslink with a unique direction (see figure 6). (A reversely ordered *Off/Set*-pair would of course reverse the direction of the crosslink.) The resulting structure can be represented by a program.

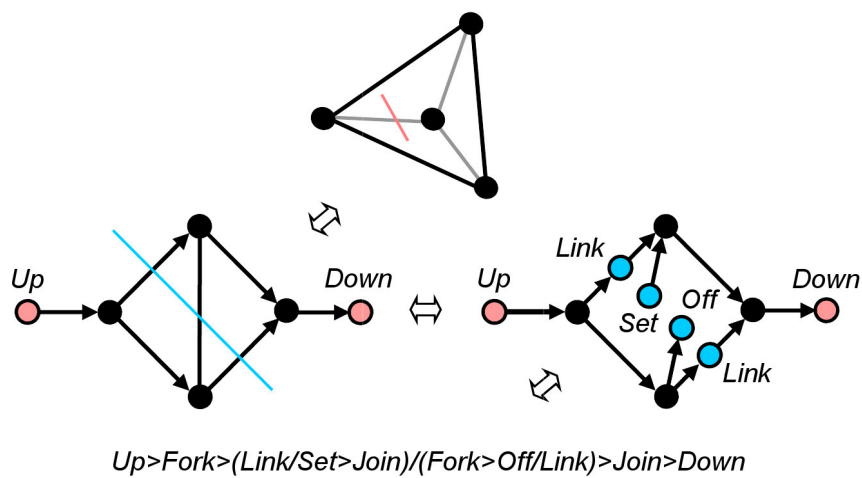

*Up>Fork>(Link/Set>Join)/(Fork>Off/Link)>Join>Down*

Figure 6: Mapping steps of a tetrahedron from the spatial structure to a planar representation, then to a fully oriented representation, and finally to a linear representation, i.e. a program.

### 5.2 Helix and Sheet
The next two examples have been selected in order to indicate the relation between *AA* and the as yet unknown protein programming language, which programs the spatial structure of proteins by chains of amino acids [1]. Since the vocabulary of *AA* is just derived from a few general principles it can be conjectured that the amino acids are also describable by *AA*-expressions. The structure (a) of figure 7 represents a model of two loops of a right-handed α-helix. Since an α-helix is a spatial structure, it takes *Up/Down*-pairs to planarize it and *Set/Off*-pairs to fully orient it from left to right. The structure (b) of figure 7 represents a model of ß-pleated sheet. Since this structure is planar, it only takes *Set/Off*-pairs in order to stretch it into a programming code.

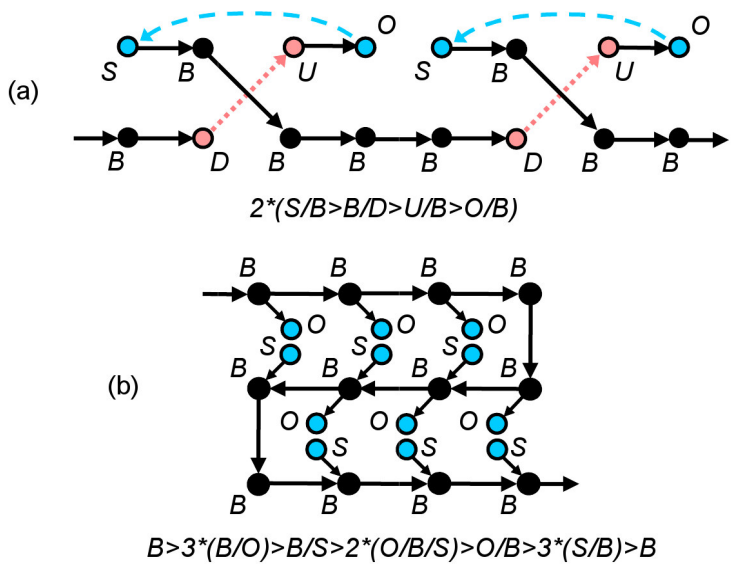

Figure 7: Model and program of a right-handed α-helix (a), and model and program of a β-sheet (b).

### 5.3 Modelling DNA
A first particular application of twin-cuts concerns the modelling of the genetic code. As well-known, the basic information on the construction of all organisms are expressed by a 4-letter programming language, with the letters being represented by the nucleobases called Adenine (a), Guanine (*G*),

Cytosine (c) and Thymine (*T*) [5]. Here we will use the same abbreviations for the nucleotides, where each nucleobase is extended by a piece of backbone. An important feature of the nucleotides is that they are pair-wise complementary, i.e. *A* matches with *T* and *G* matches with *C*. The twin-cuts of *AA* offer exactly the same property and thus are perfectly suited to model DNA. Since *AA* distinguishes between above and below, there are two representations for each of the four nucleotides. This is visualized by figure 8, where the four twin-cuts of figure 5 are now described by akton terms. Assuming that *A* is represented by the akton term *BU>BO* (*OB>UB*) then *T* is represented by *SB>DB* (*BD>BS*). Likewise, assuming that *G* is represented by *BD>BO* (*OB>DB*) then *C* is represented by *SB>UB* (*BU>BS*). Figure 8 models the four elements of DNA-code. The elements are pairwise complementary to each other. Exchanging *Up* and *Down* as well as *Set* and *Off* turns term (a) into term (b), and term (c) into term (d). That is exactly the matching property of Adenine/Thymine-pair and the Guanine/Cytosine-pair.

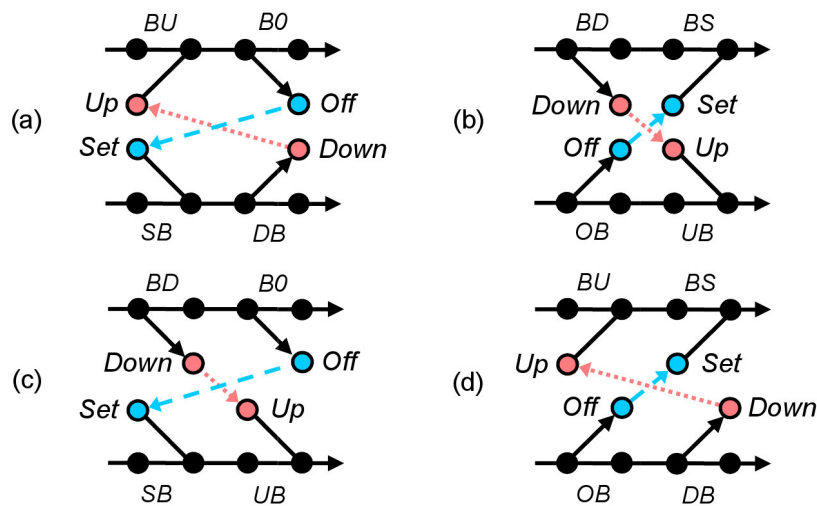

Figure 8: Models of the four elements of DNA-code.

Another particular feature of the twin-cuts is that their left- or right-twist is suited to model the chirality of a double helix. This statement is substantiated by the following arguments: Since topological nodes have a finite albeit unknown volume and twin-cuts are crossings of two pairs of nodes, twin-cuts do have a natural skew. As shown in figure 9, a left-twisted twin-cut causes a right-twisted skew between the strands and vice versa. It is this skew that causes two strands which are interconnected by twin-cuts to turn into a helix.

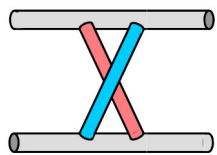

Figure 9: Simplified sketch of a twin-cut visualizing its skew.

Another example of a twin-cut structure is the circuit diagram of an *SR*-Flipflop (figure 11) which will be treated in the next section.

## 6. Concretizing AA: Symbolic Systems

The language of *AA*, as defined up to this point, describes the topological structure of abstract discrete spatial systems. It can now step by step be concretized towards special discrete systems by introducing new aktons as subsorts of the abstract akton sorts. The new subsorts inherit the properties of their hierarchical ancestors and may be provided with additional properties. However, none of the additional properties may ever conflict with the inherited properties. There are three main ways how to concretize abstract *AA*. Most trivially, it could be done by introducing new subsorts and designating them only symbolically, i.e. without adding new properties. Typical examples are the diagrams of analogue or digital circuitry. Next, the new akton sorts could be provided with functionality giving rise to functional systems, and finally they could be provided with a metric giving rise to concrete spatial systems. Each of the three ways will be studied in the sequel.

**6.1 Digital Circuit Description**
As well-known, every binary function can be realized by a single sort *Nand* or by a single sort *Nor*. Usually however, several sorts are applied. Here, we introduce the subsorts *And*, *Or*, *Not* and *Wire*, where *Not* denotes the inversion function and *Wire* the *1*-function. *And* and *Or* are subsorts of *Join*, and *Not* and *Wire* are subsorts of *Link*. The extension of the Akton hierarchy by concrete subsorts is depicted in figure 10(a).

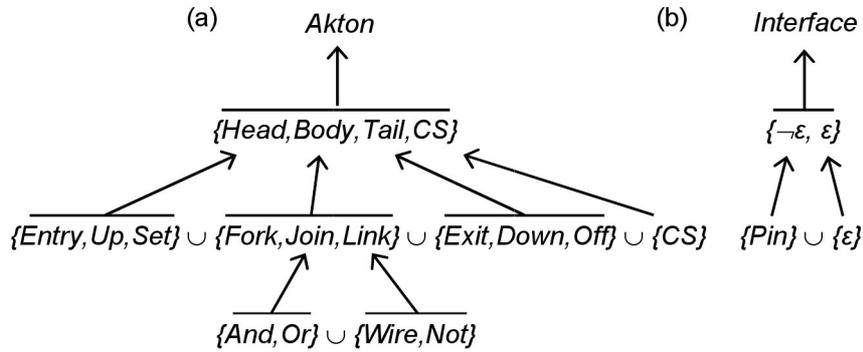

Figure 10: Extending sort Join of the akton hierarchy (a) by the subsorts And and Or, and sort Link by the subsorts Wire and Not, keeping the interface hierarchy (b) unchanged turns AA into a digital circuit description language.

The set of digital akton sorts is designated by *dA*.

**Definition D.1 (Digital Akton Sorts)**
*The set dA of digital akton sorts is defined as:*
$dA := \{Entry, Up, Set\} \cup \{Fork\} \cup Join \cup Link \cup \{Exit, Down, Off\} \cup \{CS\}$, where
$Join := \{And, Or\}$,
$Link := \{Wire, Not\}$

**Definition D.2 (Digital Akton Terms)**
*The set of digital akton terms $dA^+$ is inductively defined as the smallest set satisfying:*
- $dA^+ \subset A^+$,
- $dA \in dA^+$
- $\forall x,y \in dA^+: (x>y) \in dA^+$
- $\forall x,y \in dA^+: (x/y) \in dA^+$

The digital functions *in, out*: $dA^+ \rightarrow I^*$ are defined as $A^+ \rightarrow I^*$

The properties of the digital circuit description language defined on the set of digital akton terms $dA^+$ are demonstrated by three digital circuits and their description by an *AA*-program.

**6.2 SR-Flipflop**
The *SR*-Flipflop shown in figure 11 serves to emphasize the important property of *AA* to analytically describe feedback circuits. With this property *AA* overcomes the severe restriction of register-transfer-level programming languages which only describe the logical expressions between two storage cycles [7].

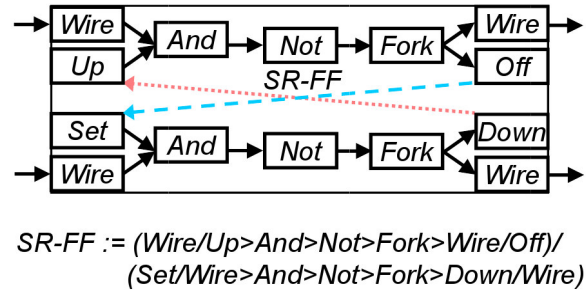

SR-FF := (Wire/Up>And>Not>Fork>Wire/Off)/
(Set/Wire>And>Not>Fork>Down/Wire)

Figure 11: Circuit diagram and AA-program of an SR-Flipflop.

**6.3 Half- and Full-Adder**
The examples of a half- and a full-adder shown in figure 12 serve to demonstrate how more complex systems can be built up from low level systems by introducing abbreviations or by concealing. Systems of arbitrary complexity can thus be treated just as every simple system.

The structure and the akton term on top of figure12 show the circuit diagram of the half-adder *HAd*, and those at the bottom the circuit diagram of the full adder *FAd*. Both akton terms are of sort $B^+$. Both diagrams are shown in order to demonstrate how complexity can be diminished by identifying akton terms by names, as *F2* for a two-Fork structure, *HAd* for a half-adder and *FAd* for a full-adder.

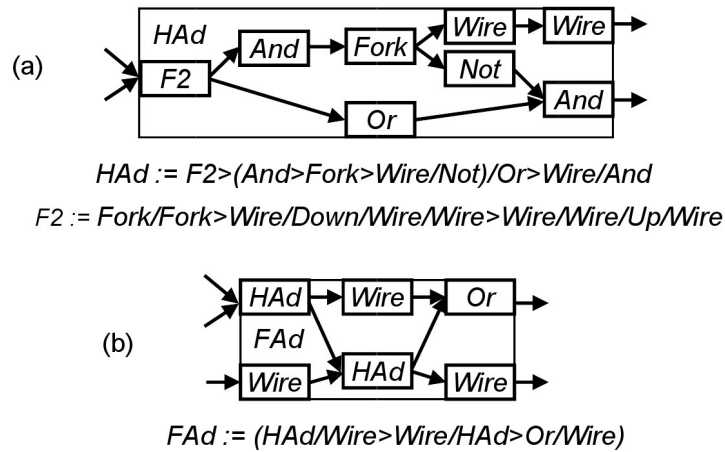

HAd := F2>(And>Fork>Wire/Not)/Or>Wire/And
F2 := Fork/Fork>Wire/Down/Wire/Wire>Wire/Wire/Up/Wire

FAd := (HAd/Wire>Wire/HAd>Or/Wire)

Figure 12. Circuit diagrams and *AA*-programs of a half-adder *HAd* (a) and a full-adder *FAd* (b).

## 7. Concretizing *AA*: Functional Systems

In the previous section we concretized *AA* symbolically, i.e. only by extending the akton hierarchy by additional aktons sorts, however without providing them with any extra properties. This step alone was sufficient to create a programming language for symbolic system description and design.

We now proceed to provide the newly introduced akton sorts with functional properties. This enhances *AA* to a general data processing language. The enhancement is achieved by introducing data values as subsorts of the interface sort *Pin* and by defining functions between the input and the output of the aktons. This way several kinds of functionality can be implemented, e.g. digital or analogue functions or even both. Moreover, because *AA* can be equipped with all the elementary functions of analogue or digital circuitry and because these functions can arbitrarily be composed to more complex functions, a plethora of low or high level programming languages could be designed.

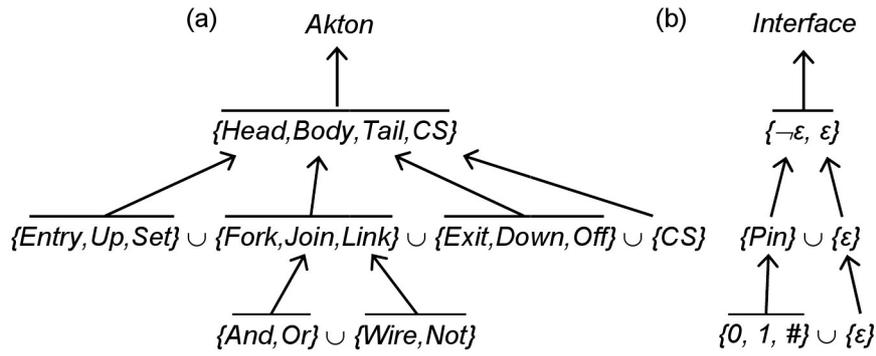

Figure 13: Extending sort *Join* by the subsorts *And* and *Or*, and sort *Link* by the subsorts *Wire* and *Not*, turns *AA* into a digital circuit description language (a). Further extending sort *Pin* of the interface (b) by the subsorts *0,1,#* and defining the functions *in* and *out* by them, as done by definition D.3, generates a digital data processing language.

It should also be noted that abstract *AA* does not impose any restrictions on the system behaviour. Since abstract *AA* does not have any states, the execution of an *AA*-program behaves flow-controlled or, if digital data processing is introduced, as data driven [6]. However, the data driven behaviour can easily be turned into a state driven one by starting the evaluation of a succeeding akton only after the output of the preceding aktons is fully defined. A clock driven behaviour, i.e. the behaviour of most computers, can then be achieved by supplying each akton with a storage function and by fitting the akton evaluation into a clock cycle.

**7.1 Digital Data Processing**
Here, we concentrate on data driven digital data processing. To this end, we extend the interface hierarchy by providing the sort *Pin* with a set of digital values *{0, 1, #}*, where # means undefined (see figure 13(b)). The value # indicates the state of the *Pin* before processing. Each of the functions contained in an akton can only be evaluated, if its individual input data are available. Moreover, since the evaluation of the functions takes a different finite amount of time, the output of an akton is always delayed and if there are several output *Pins*, they are never exactly synchronized.

**Definition D.3 (Digital Akton Interfaces)**
*The set of digital akton interfaces dI is defined as:*
$dI := Pin \cup \{\varepsilon\}$, where
$Pin := \{0, 1, \#\}$

**Definition D.4 (Digital Interface Terms)**
*The set of digital interface terms $dI^*$ is inductively defined as the smallest set satisfying:*
- $dI^* \subset I^*$,
- $dI \subset dI^*$,
- $\forall x,y \in dI^*: (x/y) \in dI^*$

**Definition D.5 (Digital Functions *in*, *out*)**
*The functions in, out: $dA^+ \rightarrow dI^*$ are defined as:*

$$\begin{aligned}
&out(And) := 1, &&\text{if } in(And)=1/1 \\
&out(And) := 0, &&\text{if } (in(And)=0/x \text{ or } in(And)=x/0) \\
&out(And) := \#, &&\text{if } (in(And)=\#/x \text{ or } in(And)=x/\#) \\
&out(Or) := 0, &&\text{if } in(Or)=0/0 \\
&out(Or) := 1, &&\text{if } (in(Or)=1/x \text{ or } in(Or)=x/1) \\
&out(Or) := \#, &&\text{if } (in(Or)=\#/x \text{ or } in(Or)=x/\#) \\
&out(Not) := 1, &&\text{if } in(Not)=0
\end{aligned}$$

$$out(Not):=0, \quad \text{if } in(Not)=1$$
$$out(Not):=\#, \quad \text{if } in(Not)=\#$$
$$out(Wire):= x, \quad \text{if } in(Wire)=x$$
$$out(Fork):=x/x, \quad \text{if } in(Fork)=x$$
$$out(Head):=x, \quad \text{if } in(Head)=\varepsilon$$
$$out(Tail):=\varepsilon, \quad \text{if } in(Tail)=x \quad\quad x \in \{1,0,\#\}$$

Establishing the transmission of data between *Next*-related akton terms extends *AA* from a description language of static systems to a description language of dynamic systems. In particular, this renders it possible to describe positive or negative feedback, i.e. the storage of data or the repetition of functions. The following definition can therefore be regarded as a basic rule of computation.

**Definition D.6 (Digital Data Transmission)**
*The digital data transmission is defined as:*
- $\forall x,y \in dA^+: in(y):=out(x), \text{ if } (x>y)$

### 7.2 Systems behaviour

The behaviour of linear *AA*-programs like those of the half-adder or the full-adder is immediately understandable just by providing them with *Entry*-data and then tracing their execution step-by-step. However, the behaviour of a cyclic program is not that simple. For this reason we inspect the behaviour of a feedback cycle as depicted in figure 14. Part (a) shows the structure of the feedback cycle and its program, part (b) the behaviour. Initially, every akton is in the undefined state #. Supplying state *0* to the input results in a steady state, where all aktons are defined. Recall that according to Def. **A.**12 *Off* transfers its input to the output of *Set*. If subsequently state 1 is supplied to the input the feedback cycle starts oscillating.

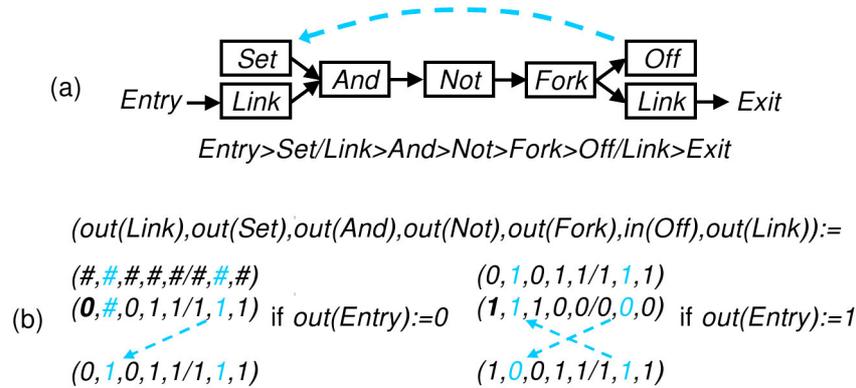

Figure 14: Structural and behavioural representation of a feedback circuit. Part (a) depicts the planar and linear representation of the circuit. Part (b) describes the states of the circuit components. In order to enter into a continuous loop the circuit first needs to be initialized by *out(Entry):=0* followed by *out(Entry):=1*.

## 8. Concretizing AA: Metrical Systems

In this chapter we are providing abstract *AA* with a metric. The metric serves to describe the shape and the size of physical systems. For this purpose a metric space is introduced as a frame of reference. Introducing a metric space requires the determination of three kinds of properties. The first property is the dimensionality of the metric space; the other two are the uniform gauge for distances and the uniform gauge for angles.

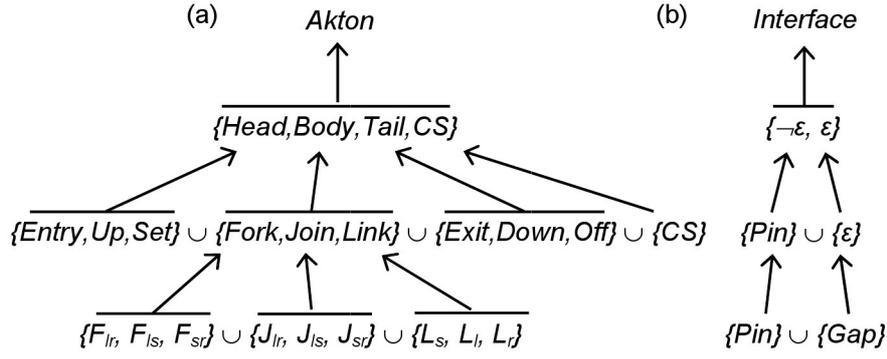

Figure 15: Extending the sorts *Fork*, *Join* and *Link* by structured metric subsorts (a) and introducing *Gap* as an identically sized companion of *Pin* (b) turns *AA* into a metric constructive programming language.

The dimensionality can be expressed by the directions a wayfarer would make use of to get from one point to any other point within the frame of reference. Three directions are needed for a planar frame, i.e. *straight*, *left* and *right*, and two more for a spatial frame, i.e. *up* and *down*.

Since there is no difference between a planar and a spatial frame of reference except for the number of directions, we confine ourselves to a planar one. The directions are introduced into *AA* by means of structured subsorts of the sorts *Fork*, *Join* and *Link*. There are three basic aktons to each subsort, i.e. $\{F_{lr}, F_{ls}, F_{sr}\}$, $\{J_{lr}, J_{ls}, J_{sr}\}$, $\{L_s, L_l, L_r\}$, where *F*, *J*, *L* are abbreviations of *Fork*, *Join* and *Link*, and the subscripts *s*, *l*, *r* are abbreviations of *straight*, *left* and *right*. (In a 3-dimensional reference frame the number of basic aktons would rise to 10 for subsorts of *Fork* and *Join* and to 5 for subsorts of *Link*.) The extended hierarchy of akton sorts is shown in figure 15(a).

**8.1 Rectangular metric Systems**

A uniform metric gauge is introduced by extending the interface of sort ε by a subsort *Gap* as shown in figure 15(b). Sort *Gap* represents an empty place with the same breadth as sort *Pin*. A uniform angular gauge is introduced by assuming rectangularity.

**Definition M.1 (Metric Akton Sorts)**
*The set of metric akton sorts mA is defined as:*
$mA := \{Entry, Up, Set\} \cup Fork \cup Join \cup Link \cup \{Exit, Down, Off\}$, where
$$Fork := \{F_{lr}, F_{ls}, F_{sr}\},$$
$$Join := \{J_{lr}, J_{ls}, J_{sr}\},$$
$$Link := \{L_s, L_l, L_r\}.$$

As defined in **A.4**, an abstract basic interface is either empty or contains *Pins*. With the metric refinement an interface of an akton term may now contain *Pins*, *Gaps* or both.

**Definition M.2 (Metric Akton Interfaces)**
*The set of metric akton interfaces mI is defined as:*
$mI := \{Pin\} \cup \varepsilon$, where
$$\varepsilon := \{Gap\}$$

**Definition M.3 (Metric Interface Terms)**
*The set of metric interface terms $mI^+$ is inductively defined as the smallest set satisfying:*
- $mI^+ \subset I^*$
- $mI \subset mI^+$
- $\forall x,y \in mI^+: (x/y) \in mI^+$

On the other hand, by introducing a metric each akton term now attains an individual size. Even the adjacent sides of two *Next*-related akton terms $x>y$ may differ in size, because of different numbers of *Gaps* above and below their proper interfaces. These *Gaps* can formally be removed by introducing two functions, called *atrim* and *btrim*, where *atrim* eliminates all *Gaps* above a first *Pin* and *btrim* all *Gaps* below a last *Pin*. The function composition of *atrim* and *btrim*, i.e. the function *trim*, produces an interface beginning and ending with a *Pin*. This interface can be interpreted as a plug. Plugging the interfaces of two *Next*-related terms $x>y$ then means to shift both terms into their correct relative *Juxta*-position.

**Definition M.4 (Functions *atrim*, *btrim*, *trim*)**
The functions *atrim*, *btrim*, *trim*: $mI^+ \rightarrow mI^+$ are inductively defined as:
$$atrim(i) := atrim(j) \text{ while } i=Gap/j,$$
$$btrim(i) := btrim(j) \text{ while } i=j/Gap$$
$$trim := atrim \circ btrim.$$

The functions $s_i$ extract the metric interfaces of the four sides of a rectangular akton term. The sides are clockwise enumerated, $s_0$ denoting the left-hand side. We first define the sides of the basic aktons.

**Definition M.5 (Functions $s_i$)**
The functions $s_i: mA^+ \rightarrow \{side_i\}$, $side_i \in mI^+$, $i \in \{0,1,2,3\}$ are defined as:

$$s_i(z) := \begin{cases} \{Gap,Gap,Pin,Gap\}, & \text{if } z=Head \\ \{Pin,Gap,Gap,Gap\}, & \text{if } z=Tail \\ \{Gap,Gap,Gap,Gap\}, & \text{if } z=CS \\ \{Pin,Pin,Gap,Pin\}, & \text{if } z=F_{lr} \\ \{Pin,Pin,Pin,Gap\}, & \text{if } z=F_{ls} \\ \{Pin,Gap,Pin,Pin\}, & \text{if } z=F_{sr} \end{cases} \quad s_i(z) := \begin{cases} \{Gap,Pin,Pin,Pin\}, & \text{if } z=J_{lr} \\ \{Pin,Pin,Pin,Gap\}, & \text{if } z=J_{ls} \\ \{Pin,Gap,Pin,Pin\}, & \text{if } z=J_{sr} \\ \{Pin,Gap,Pin,Gap\}, & \text{if } z=L_s \\ \{Pin,Pin,Gap,Gap\}, & \text{if } z=L_l \\ \{Pin,Gap,Gap,Pin\}, & \text{if } z=L_r \end{cases}$$

Next we define the structural relations between the output and the input of each metric subsort.

**Definition M.6 (Metric Functions *in*, *out*)**
The metric functions *in*, *out*: $mA^+ \rightarrow mI^+$ are defined as:
$$out(L_s) := s_2, \text{ if } in(L_s) = s_0$$
$$out(L_l) := s_1, \text{ if } in(L_l) = s_0$$
$$out(L_r) := s_3, \text{ if } in(L_r) = s_0$$
$$out(F_{ls}) := s_1/s_2, \text{ if } in(F_{ls}) = s_0$$
$$out(F_{sr}) := s_2/s_3, \text{ if } in(F_{sr}) = s_0$$
$$out(F_{lr}) := s_1/s_3, \text{ if } in(F_{lr}) = s_0$$
$$out(J_{ls}) := s_2, \text{ if } in(J_{ls}) = s_1/s_0$$
$$out(J_{sr}) := s_2, \text{ if } in(J_{sr}) = s_0/s_3$$
$$out(J_{lr}) := s_2, \text{ if } in(J_{lr}) = s_1/s_3$$

**Definition M.7 (Metric Akton Terms)**
The set of metric akton terms $mA^+$ is inductively defined as the smallest set satisfying:
- $mA^+ \subset A^+$
- $mA \subset mA^+$
- $\forall x,y \in mA^+: (x>y) \in mA^+$, if $trim(out(x)) = trim(in(y))$

In order to fold a network of rigid rectangular aktons into a planar area, we need a means to turn the akton terms into another direction. This can be achieved by introducing two tilt functions *tl* and *tr*, where *tl* turns an akton term orthogonally to the left and *tr* orthogonally to the right.

**Definition M.8 (Tilt Functions tl (anticlockwise), tr (clockwise))**
*The tilt functions tl, tr: $mA^+ \to mA^+$ are defined as:*
$$t(x{>}y) = t(x){>}t(y), \quad t(x/y) = t(x)/t(y), \quad t \in \{tl, tr\},$$
$$tl(tr(x)) = tr(tl(x)) = x,$$
$$tl(tl(x)) = tr(tr(x))$$

Assigning input/output directions to the *Body* aktons extends them to three subsorts each. The basic sorts of structured metric aktons are depicted in figure 16.

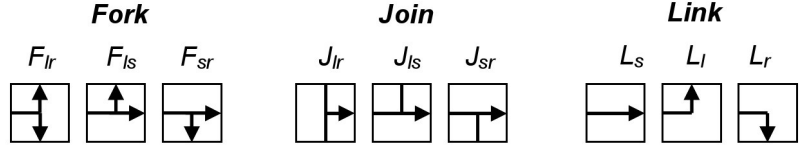

Figure 16: Basic sorts of structured metric aktons. There are three subsorts of *Fork*, *Join* and *Link*, which differ in the direction they proceed. Their paths may turn either to the left, to the right or proceed straight.

Abstract *Multiple Links*, *multiple Forks* and *multiple Joins* have already been defined by Defs. A.18, A.19 and A.20. These structures are important in the layout process of digital systems. According to the rectangular metric being assumed here, there are three structures of *Multiple Links*, which by insertion or deletion serve to define the desired folding. They are based on the metric subsorts $L_s$, $L_l$, $L_r$. While a straight *multiple Link* can just be generated by *Juxta*-relating *Links* to columns and then *Next*-relating the columns to strips of any finite length, tilted multiple structures need to be realized by *Juxta*-related structures of single chains of ascending and descending length which together form a square. The centerpiece of a left-tilted chain is an element of sort $L_l$ and the centerpiece of a right-tilted chain is an element of sort $L_r$.

**Definitions M.9 (Metric Multiple Links)**
*The sets of metric multiple Links are recursively defined as:*
- $mL^+ \subset mA^+$
- $mL_s^+ \cup mL_l^+ \cup mL_r^+ \subset mL^+$
- $x(i) \in mL_s^+$, if $x(0)=L_s$, else $x(i):=x(i-1)/x(i)$, if $i>0$
- $x(i) \in mL_l^+$, if $x(0)=L_l$, else $x(i):=x(i-1)/(i*L_s{>}L_l{>}tl(i*L_s))$, if $i>0$
- $x(i) \in mL_l^+$, if $x(0)=L_r$, else $x(i):=(i*L_s{>}L_r{>}tr(i*L_s))/x(i-1)$, if $i>0$

Metric *multiple Fork* and *multiple Join* structures both exhibit some special features. Firstly, there is no *multiple Fork* of sort $F_{lr}$ (or *Join* of sort $J_{lr}$) because the two outputs of such a structure are ordered reversely. Secondly, there is only one structure of a *multiple Fork* of sort $F_{ls}$ (or *Join* of sort $J_{ls}$) and only one of a multiple *Fork* of sort $F_{sr}$ (or *Join* of sort $J_{sr}$). The reason for the second restriction is that by definition *Down*- and *Up*-terms are always separated by a *B*-term (see figure 2 and Def. A.9). This restriction is only met if the *Down/Up*-cut is realized by an *l*- or an *r*-leg of a multiple *Fork* of sort $F_{ls}$ or $F_{sr}$ or a multiple *Join* of sort $J_{ls}$ or $J_{sr}$ (see figure 16). Because the *Down*- and *Up*-terms may be alternatively at the left or at the right side of a *B*-term there are two different *Fork* junctions and two different *Join* junctions. For clarity we conceal these junctions into aktons calling them $F_{ld}$, $F_{rd}$, $J_{lu}$, $J_{ru}$.
- $F_{ld}:=conceal(F_{ls}{>}tl(Down)/(L_l{>}tl(L_s))$, $F_{ld} \in DB^+$
- $F_{rd}:=conceal(F_{sr}{>}(L_r{>}tr(L_s))/tr(Down))$, $F_{rd} \in BD^+$
- $J_{lu}:=conceal(tr(Up/(L_s{>}L_l)){>}J_{ls})$, $J_{lu} \in UB^+$
- $J_{ru}:=conceal(tl((L_s{>}L_r)/Up){>}J_{sr})$, $J_{ru} \in BU^+$

A concealed structure can be scaled down into a smaller rectangle if there are *Gaps* on opposite sides. The only effect is that the original structure cannot explicitly be represented any more. In order to further simplify the subsequent definitions we shrink the junctions to the least size of a metric akton, i.e. unit square size. The shrinkage reduces the original term structures to a tilted *Link* and an undirected via each, as shown in figure 17. The sides of the junctions are identified by $s_i(z)$ as defined in Def. M.5.

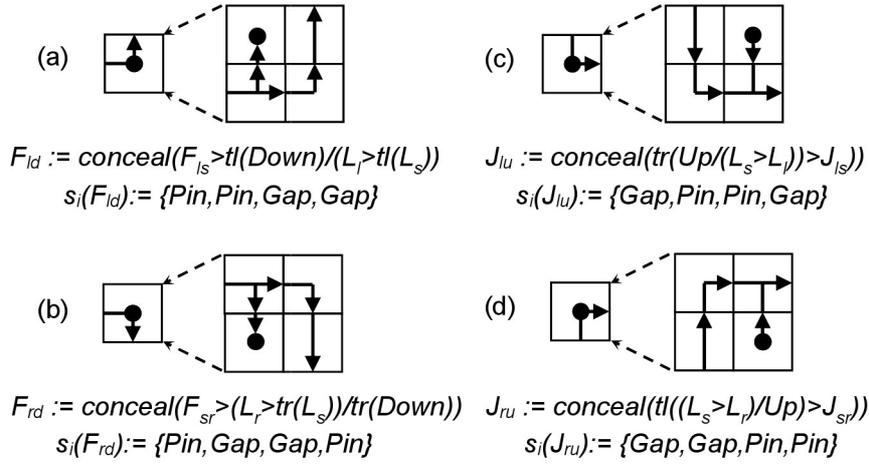

Figure 17: A concealed structure can be scaled down into a smaller rectangle if there are Gaps on opposite sides. Concerning the structure of *multiple Forks* and *multiple Joins* treated here, they can be reduced to unit square which only shows a tilted Link and a via.

**Definitions M.10 (Metric Multiple Forks)**
*The sets of metric multiple Forks are recursively defined as:*
- $mF^+ \subset mA^+$
- $mF_{ls}^+ \cup mF_{sr}^+ \subset mF^+$
- $x(i) \in mF_{ls}^+$,
  $pre(x(i)):= F_{ld}$, $suc(x(i)):=tl(L_s/CS)/Up$, *if $i=0$*
  $pre(x(i)):=pre(x(i-1))/(i*L_s>F_{ld}>tl(i*L_s))$, $suc(x(i)):=tl(L_s\wedge(i+1)/CS)/Up\wedge(i+1)$, *if $i>0$*
- $x(i) \in mF_{sr}^+$,
  $pre(x(i)):=F_{rd}$, $suc(x(i)):=Up/tr(L_s/CS))$, *if $i=0$*
  $pre(x(i)):=(i*L_s>F_{rd}>tr(i*L_s))/pre(x(i-1))$, $suc(x(i)):=Up\wedge(i+1)/tr(CS/L_s\wedge(i+1))$, *if $i>0$*

**Definitions M.11 (Metric Multiple Joins)**
*The sets of metric multiple Joins are recursively defined as:*
- $mJ^+ \subset mA^+$
- $mJ_{ls}^+ \cup mJ_{sr}^+ \subset mJ^+$
- $x(i) \in mJ_{ls}^+$,
  $pre(x(i)):=tr(L_s/CS)/Down$, $suc(x(i)):=J_{lu}$, *if $i=0$*
  $pre(x(i)):= tr(L_s\wedge(i+1)/CS)/Down\wedge(i+1)$, $suc(x(i)):=suc(x(i-1))/(tr(i*L_s)>J_{lu}>i*L_s)$, *if $i>0$*
- $x(i) \in mJ_{sr}^+$,
  $pre(x(i)):= Down/tl(CS/L_s)$, $suc(x(i)):=J_{ru}$, *if $i=0$*
  $pre(x(i)):= Down\wedge(i+1)/tl(CS/L_s\wedge(i+1))$, $suc(x(i)):=(tl(i*L_s)>J_{ru}>i*L_s)/suc(x(i-1))$, *if $i>0$*

**8.2 Layout of metric triple Links and triple Forks**
*Metric Links* and *metric Forks* are important structures for the planar layout of electronic systems, because they allow treating a bundle of parallel connections like a single one. Metric Links are particularly indispensable for folding a straight akton structure into a desired planar layout. Each time the structure is to be expanded a metric *multiple Link of subsort $mL_s$* has to be inserted, and each time it is to be tilted to the left or to the right either subsort $mL_l$ or $mL_r$ has to be inserted. *Metric Links* and *metric Forks* are important structures for the planar layout of electronic systems, because they allow treating a bundle of parallel connections like a single one. That is the reason why we select them as examples. Figure 18 (a) shows the structure and the program of a left-tilted strip of three chains of

*Links*. Figure 18 (b) shows the structure and the program of a *multiple Fork* of three strips. Sort $F_{ld}$ represents a unit square concealment of the term $(F_{ls}>tl(Down)/(L_l>tl(L_s))$.

Please observe that the structure inside the unit square of an elementary akton is not constituent of the metric introduced by $mA^+$. The *metric Links* and *Forks* may therefore be provided with a finer structure, as done here. Copper-colored regions are meant to depict conducting areas while white regions are meant to depict insulated areas.

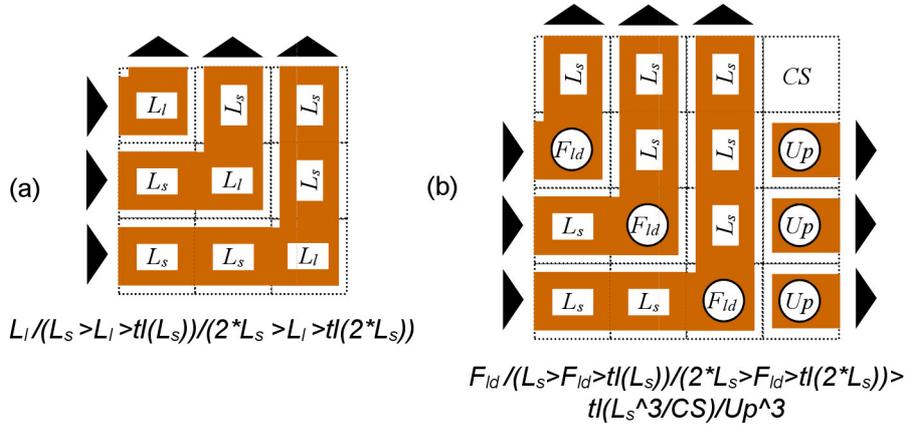

Figure 18: Two layout examples: Structure and program of a strip of three chains turning left (a), and of a strip of three Forks forking straight and left (b).

## 9. Conclusion

This paper comprehends 4 major findings:
1. Every discrete physical system reduces to a spatiotemporal topological network of nodes, if the functional and metric properties are abstracted.
2. There exists a homeomorphism, i.e. a bijective and bicontinuous mapping, between a spatiotemporal topological network and an executable sequence of symbols, i.e. a program written in a topological programming language. Total execution of the topological program reconstructs the original spatiotemporal network; execution without healing the spatial cuts produces a planar version.
3. Reintroducing the functionality of the physical system turns the abstract programming language into a flow-controlled general data processing language. The functions may be analogue or digital. The flow-control can be restricted by partial synchronization or by total clock-control. It is then possible to squeeze the digital data processing into the word-at-time scheme of the von-Neumann-architecture, as most digital programming languages of today.
4. Reintroducing the metrics of the physical system, i.e. the shape and size of the components, turns abstract Akton-Algebra into a novel hardware system construction language. The language is particularly suited for the layout of complex digital systems.

As yet Akton-Algebra has only been realized as a basic programming language by which the programming of physical structures like those of biomolecules or circuit diagrams can be demonstrated.

## Acknowledgments

There are several colleagues who notably supported me during my long work towards Akton-Algebra. One of the first was Rudolf Albrecht, Innsbruck, who showed great interest in my approach when we first met in 1993 in Lessach and with whom I have had many very fruitful discussions since then. A next one to be mentioned gratefully is Walter Dosch, Lübeck, for several lucid comments. And finally, I am most obliged and grateful to Bernd Braßel, Kiel, with whom I have been in an ongoing email discussion for about four years. Again and again pointing his finger to mistakes and flaws, he taught me to turn my intuitive ideas into logical statements.